\documentclass[aps,pra,showpacs,amssymb,superscriptaddress,nofootinbib,twocolumn]{revtex4}

\usepackage{graphicx}

\begin{document}

\title{Dynamics of momentum entanglement in lowest-order QED}

\author{L. Lamata} \email{lamata@imaff.cfmac.csic.es}

\affiliation{Instituto de Matem\'aticas y F\'{\i}sica Fundamental,
CSIC, Serrano 113-bis, 28006 Madrid, Spain}

\affiliation{Max-Planck-Institut f\"{u}r Quantenoptik,
Hans-Kopfermann-Strasse 1, D-85748 Garching, Germany}

\author{J. Le\'on} \email{leon@imaff.cfmac.csic.es}

\affiliation{Instituto de Matem\'aticas y F\'{\i}sica Fundamental,
CSIC, Serrano 113-bis, 28006 Madrid, Spain}

\author{E. Solano} \email{enrique.solano@mpq.mpg.de}

\affiliation{Max-Planck-Institut f\"{u}r Quantenoptik,
Hans-Kopfermann-Strasse 1, D-85748 Garching, Germany}

\affiliation{Secci\'on F\'{\i}sica, Departamento de Ciencias,
Pontificia Universidad Cat\'olica del Per\'u, Apartado Postal
1761, Lima, Peru}

\begin{abstract}
We study the dynamics of momentum entanglement generated in the
lowest-order QED interaction between two massive spin-1/2 charged
particles, which grows in time as the two fermions exchange
virtual photons. We observe that the degree of generated
entanglement between interacting particles with initial
well-defined momentum can be infinite. We explain this divergence
in the context of entanglement theory for continuous variables,
and show how to circumvent this apparent paradox. Finally, we
discuss two different possibilities of transforming momentum into
spin entanglement, through dynamical operations or through Lorentz
boosts.
\end{abstract}


 \maketitle

\section{Introduction\label{s1}}

Entanglement is the physical property that Schr\"{o}dinger
described as \textit{Not one but the characteristic trait of
quantum mechanics}. It has played a fundamental role in the study
of the completeness of quantum mechanics \cite{epr,Bell64}.
Nowadays, in quantum information theory, entanglement is
considered as a physical resource, equivalent in many aspects to
the role energy played in classical and quantum mechanics. Most
applications in this novel field, like quantum
teleportation~\cite{Bennett93}, quantum
communication~\cite{QuantumComm}, quantum
cryptography~\cite{Ekert91}, and some algorithms of quantum
computation are carried out by using this intriguing quantum
property~\cite{NielsenChuang}. A thorough study of entanglement in
quantum information theory would demand a natural classification
between discrete~\cite{MiguelAngel} and continuous
variables~\cite{Englert}.

In the last few years two apparently different fields,
entanglement and relativity, have experienced intense research in
an effort for treating them in a common
framework~\cite{C97,PST02,AM02,GA02,GBA03,PS03,TU03,ALM03,PT04,MY04}.
Most of those works investigated the Lorentz covariance of
entanglement through purely kinematic considerations, and only a
few of them studied {\it ab initio} the entanglement dynamics. For
example, in the context of Quantum Electrodynamics (QED), Pachos
and Solano \cite{PS03} considered the generation and degree of
entanglement of spin correlations in the scattering process of a
pair of massive spin-$1/2$ charged particles, for an initially
pure product state, in the low-energy limit and to the lowest
order in QED. Manoukian and Yongram \cite{MY04} computed the
effect of spin polarization on correlations in a similar model,
but also for the case of two photons created after $e^+e^-$
annihilation, analyzing the violation of Bell's inequality
\cite{Bell64}. In an earlier work, Grobe et al. \cite{qedentang}
studied, in the nonrelativistic limit, the dynamics of
entanglement in position/momentum of two electrons which interact
with each other and with a nucleus via a smoothed Coulomb
potential. They found that the associated quantum correlations
manifest a tendency to increase as a function of the interaction
time.

In this paper, we study to the lowest order in QED the interaction
of a pair of identical, charged, massive spin-1/2 particles, and
how this interaction increases the entanglement in the particle
momenta as a function of time. We chose to work at lowest order,
where entanglement already appears full-fledged, precisely for its
simplicity. In particular this allows to set aside neatly other
intricacies of QED, whose influence on entanglement should be
subject of separate analysis. Here, the generation of entanglement
is a consequence of a conservation law: the total relativistic
four-momentum is preserved in the system evolution. This will also
be the case in any interaction verifying this conservation law, as
occurs in closed multipartite systems, while allowing the change
in the individual momentum of each component. In the asymptotic
limit, the infinite spacetime intervals involved in the S-matrix
result in the generation of an infinite amount of entanglement for
interacting particles with well-defined momentum. QED is a place
where infinities can be avoided, and this will be also true, even
though for other physical reasons, in the case of divergences
appearing in momentum entanglement, a distinctive feature of
continuous variables~\cite{Englert}. We will also discuss two
different possibilities of establishing transfer of entanglement
between momentum and spin degrees of freedom in the collective
two-particle system: through dynamical operations or Lorentz
boosts.

In Sec. \ref{fte}, we analize at lowest order and at finite time the
generation of momentum entanglement between two electrons. In Sec.
\ref{ges}, we calculate the Schmidt decomposition of the amplitude
of a pair of spin-1/2 particles, showing the growth of momentum
entanglement as they interact via QED. We obtain also analytic
approximations of the Schmidt modes, both, in momentum and
configuration spaces. In Appendix \ref{entangfi}, we include some
notations and definitions related to entanglement theory for
discrete and continuous variables. In Appendix \ref{maj}, we address
the possibilities of transferring entanglement between momenta and
spins via dynamical action, with Local Operations and Classical
Communication (LOCC), or via kinematical action, with Lorentz
transformations.

\section{Two electron Green function in perturbation theory\label{fte}}

To address the properties of entanglement of a two electron system
one needs the amplitude (wave function) $\psi(x_1,x_2)$ of the
system, an object with 16 spinor components dependent on the
configuration space variables $x_1$, $x_2$ of both particles. The
wave functions were studied perturbatively by  Bethe and
Salpeter~\cite{bethe} and their evolution equation was also given
by Gell-Mann and Low~\cite{gellmann}. The wave function
development is closely related to the two particle Green function,
\begin{equation}
K(1,2;3,4)\, = \, (\Psi_0, T(\psi(x_1)\psi(x_2)\bar{\psi}(x_3)
\bar{\psi}(x_4))\,\Psi_0)\label{m1}
\end{equation}
which describes (in the Heisenberg picture) the symmetrized
probability amplitude for one electron to proceed from the event
$x_3$ to the event $x_1$ while the other proceeds from $x_4$ to
$x_2$. If $u_{\mathbf{p}s}(3)$ describes the electron at 3 and
$u_{\mathbf{p}'s'}(4)$ the one at 4, then
\begin{eqnarray}
\psi(x_1,x_2)&=&\int d\sigma_\mu(3)\,d\sigma_\nu(4) \,K(1,2;3,4)
\gamma^\mu_{(3)}\,\gamma^\nu_{(4)}\nonumber\\&\times&
u_{\mathbf{p}s}(3)\, u_{\mathbf{p}'s'}(4),
\end{eqnarray}
 will be their correlated amplitude at 1,
2. In the free case this is just $u_{\mathbf{p}s}(1)\,
u_{\mathbf{p}'s'}(2)$ , but the interaction will produce a
reshuffling of momenta and spins that may lead to entanglement.
The two body Green function $K$ is precisely what we need for
analysing the dynamical generation of entanglement between both
electrons.

\begin{figure}
\begin{center}
\includegraphics[width=7.5cm]{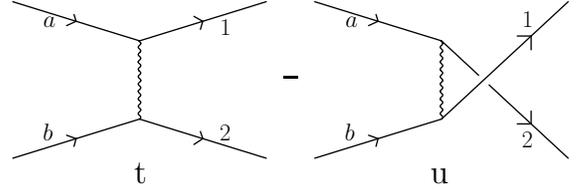}
\end{center}
\caption{Feynman diagrams for the QED interaction between two
electrons (second order). The minus sign denotes the antisymmetry of
the amplitude associated to the fermion statistics.
\label{figqed1bis}}
\end{figure}

Perturbatively~\cite{bethe},
\begin{eqnarray}
K(1,2;3,4)  \!\! & = & \!\! S_F(1,3) \, S_F(2,4)\,-\, e^2 \!\!
\int d^4x_5\,d^4x_6 S_F(1,5) \nonumber \\  & \times & \!\!\!
S_F(2,6)\,\gamma_{(5)}^{\mu} D_F(5,6) \,
{\gamma_{(6)}}_{\mu} S_F(5,3) \, S_F(6,4)\nonumber\\
&+&\cdots -\{1\,\leftrightarrow\,2\}\label{m4}
\end{eqnarray}
where all the objects appearing in the expansion are those of a
free field theory. We may call $K^{(n)}$ to the successive terms
on the right hand side of this expression. They will describe the
transfer of properties between both particles  due to the
interaction. This reshuffling vanishes at lowest order, which
gives just free propagation forward in time:
\begin{eqnarray}
\lefteqn{\int d^3 x_1 d^3 x_2 d^3 x_3 d^3 x_4
u^\dagger_{\mathbf{p_1} s_1}(1) \, u^\dagger_{\mathbf{p_2}
s_2}(2)K^{(0)}(1,2;3,4)}\nonumber\\&&\times\gamma_{(3)}^0
u_{\mathbf{p_a} s_a}(3) \,\gamma_{(4)}^0 u_{\mathbf{p_b} s_b}(4)=
\theta(t_1-t_3) \, \theta(t_2-t_4)\nonumber\\&&\times
\delta_{s_1s_a}\,\delta_{s_2 s_b} \delta^{(3)}(\mathbf{p}_1 -
\mathbf{p}_a) \,\delta^{(3)}(\mathbf{p}_2 -
\mathbf{p}_b)\label{m41}
\end{eqnarray}
\noindent where $u_{\mathbf{p} s}(x)\,=(2\pi)^{-3/2}(m/ E)^{1/2}
\exp(-ipx) u_s(\mathbf{p})$. The first effects of the interaction
appear when putting $K^{(2)}$ instead of $K^{(0)}$ in the left hand
side of the above equation. The corresponding process is shown in
Fig. \ref{figqed1bis}.
To deal with this case we choose $t_1=t_2=t, \, t_3=t_4=-t$ and
introduce the new variables
\begin{eqnarray}
\!\!\!\!\!\! t_+&=&\frac{1}{2}(t_5+t_6) , \;\;\; t_+\in(-t,t),
\label{eqfte10}\\
\!\!\!\!\!\! t_- &=&\frac{1}{2}(t_5-t_6) , \;\;\;
t_-\in(-(t-|t_+|),t-|t_+|) , \label{eqfte11}
\end{eqnarray}
in Eq.~(\ref{m4}), yielding
\begin{eqnarray}
&& \!\!\!\!\!\! \lefteqn{\tilde{K}^{(2)}(1,2;a,b;t)=} \nonumber\\
 && \!\!\!\! \frac{2ie^2}{(2\pi)^4} \frac{j^{\mu}_{1a}j_{\mu
2b}}{\sqrt{2E_12E_22E_a2E_b}} \delta^{(3)}
(\mathbf{p}_1+\mathbf{p}_2-\mathbf{p}_a-\mathbf{p}_b) \nonumber \\
&&  \!\!\!\!\!\!  \times \int_{- \! \infty}^{\infty} \!\!
\frac{dk^0}{(k^0)^2-(\mathbf{p}_a-\mathbf{p}_1)^2 + i \epsilon} \!
\int_{-t}^{t} \!\!\!\! dt_+ \mathrm{e}^{-i(E_a+E_b-E_1-E_2)t_+}
\nonumber \\ &&  \!\!\!\!\!\! \times
\int_{-(t-|t_+|)}^{t-|t_+|}dt_-
\mathrm{e}^{-i(E_1-E_2+E_b-E_a+2k^0)t_-}-\{1\leftrightarrow 2\} \,
, \label{eqfte12}
\end{eqnarray}
where $\tilde{K}^{(2)}(1,2;a,b;t)$ is a shorthand notation for what
corresponds to (\ref{m41}) at second order, and $j^{\mu}_{kl}\,=\,
\bar{u}_k\gamma^{\mu}u_l$. After some straightforward calculations,
we obtain
\begin{eqnarray}
&&\tilde{K}^{(2)}(1,2;a,b;t) = \frac{e^2}{4\pi^3}\frac{\delta^{(3)}
(\mathbf{p}_1+\mathbf{p}_2-\mathbf{p}_a-\mathbf{p}_b)}{\sqrt{2E_12E_22E_a2E_b}}
\nonumber \\ && \times \{j^{\mu}_{1a}j_{\mu
2b}[S_t(t)+\Upsilon_t(t)]-j^{\mu}_{2a}j_{\mu
1b}[S_u(t)+\Upsilon_u(t)]\},\nonumber\\
\end{eqnarray}
with,
\begin{eqnarray}
S_t(t)&=&\frac{i}{\left(\frac{E_1-E_2+E_b-E_a}{2}\right)^2
-(\mathbf{p}_a-\mathbf{p}_1)^2} \nonumber\\ && \times
\frac{\sin{[(E_1+E_2-E_a-E_b)t]}}{E_1+E_2-E_a-E_b},\label{eqfte17}\\
\Upsilon_t(t) = \!\!\!\!\! &&
\frac{1}{|\mathbf{p}_a-\mathbf{p}_1|} \nonumber \\ && \,\, \times
\left\{   i \left[ \frac{1}
{\mu(\Sigma^2-\mu^2)}+\frac{1}{\nu(\Sigma^2-\nu^2)}\right]\right.
\Sigma\sin(\Sigma t) \nonumber \\  && \,\,\,\, -
\left[\frac{1}{\Sigma^2-\mu^2}+\frac{1}{\Sigma^2-\nu^2}\right]
\cos(\Sigma t)\nonumber \\ && \,\,\,\, +
\left.\left[\frac{1}{\Sigma^2-\mu^2}\mathrm{e}^{-i\mu
t}+\frac{1}{\Sigma^2-\nu^2}\mathrm{e}^{-i\nu t}\right]\right\},
\label{eqfte17bis}
\end{eqnarray}
\begin{eqnarray}
\Sigma&=&E_1+E_2-E_a-E_b\label{eqfte17bis2},\\
\mu&=&\Delta+2|\mathbf{p}_a-\mathbf{p}_1|,\label{eqfte17bis3}\\
\nu&=&-\Delta+2|\mathbf{p}_a-\mathbf{p}_1|,\label{eqfte17bis4}\\
\Delta&=&E_1-E_2+E_b-E_a,\label{eqfte17bis5}
\end{eqnarray}
and
\begin{eqnarray}
&S_u(t)\leftrightarrow S_t(t),
\Upsilon_u(t)\leftrightarrow\Upsilon_t(t),&\nonumber\\
&1 \leftrightarrow 2&
\end{eqnarray}
$S_{t,u}$ are the only contributions that remain asymptotically
($t\rightarrow \infty$) leading to the standard scattering
amplitude, while $\Upsilon_{t,u}$ vanish in this limit. We recall
that these are weak limits: no matter how large its modulus, the
expression in Eq.~(\ref{eqfte17bis}) will vanish weakly due to its
fast oscillatory behavior. On the other hand, the sinc function in
Eq.~(\ref{eqfte17}) enforces energy conservation via
\begin{equation}
\lim_{t\rightarrow\infty}\frac{\sin{[(E_1+E_2-E_a-E_b)t]}}{E_1+E_2-E_a-E_b}
=\pi\delta(E_1+E_2-E_a-E_b) . \label{eqfte18}
\end{equation}
This limit shows also that the entanglement in energies increases
with time~\cite{lljl}, see Appendix \ref{entangfi}, reaching its
maximum (infinite) value when $t\rightarrow \infty$ for particles
with initial well-defined momenta and energy. This result is
independent of the chosen scattering configuration. Exact energy
conservation at large times, united to a sharp momentum
distribution of the initial states, would naturally result into a
high degree of entanglement. The better defined the initial
momentum of each electron, the larger the asymptotic entanglement.
The physical explanation to this unbounded growth is the
following: the particles with well defined momentum (unphysical
states) are spread over all space, and thus their interaction is
ubiquitous, with the consequent unbounded degree of generated
entanglement. This is valid for every experimental setup, except
for those pathological cases where the amplitude cancels out, due
to some symmetry. In the following section, and for illustrative
purposes, we will single out these two possibilities.

i) The case of an unbounded degree of attainable entanglement due
to an incident electron with well defined momentum. We consider,
with no loss of generality, a fuzzy distribution in momentum of
the second initial electron, for simplicity purposes.

ii) Basically the same setup as in (i) but with a specific spin
configuration, which leads to cancellation of the amplitude at
large times due to symmetry, and thus to no asymptotic
entanglement generation.

 On the other hand, for finite times, nothing prevents a sizeable
contribution from Eq.~(\ref{eqfte17bis}). In fact, in the limiting
case where $t^{-1}$ is large compared to the energies relevant in
the problem, it may give the dominant contribution to entanglement.
Whether the contribution from $\Upsilon_t(t)$ and $\Upsilon_u(t)$ is
relevant, or not, depends on the particular case considered.

\section{Two electron entanglement generation at lowest order\label{ges}}

The electrons at $x_3, \,x_4$ will be generically described by an amplitude $F$
\begin{eqnarray}
\psi_F(x_3,x_4)=\sum_{s_a,s_b}\int d^3\mathbf{p}_a \! \int \!
d^3\mathbf{p}_b\, F(\mathbf{p}_a,s_a;\mathbf{p}_b,s_b)\nonumber\\
\times
u_{\mathbf{p}_a,s_a}(x_3)\, u_{\mathbf{p}_b,s_b}(x_4)
\end{eqnarray}
that should be
normalizable to allow for a physical interpretation, i.e.,
\begin{eqnarray}
\sum_{s_a,s_b}\int d^3\mathbf{p}_a \! \int \!
d^3\mathbf{p}_b|F(\mathbf{p}_a,s_a;\mathbf{p}_b,s_b)|^2=1.
\end{eqnarray}
For separable states where
$F(a;b)=f_a(\mathbf{p}_a,s_a)f_b(\mathbf{p}_b,s_b)$,  $f_a$ and
$f_b$ could be Gaussian amplitudes $g$ centered around a certain
fixed momentum $\mathbf{p}^0$ and a certain spin component~$s^0$,
\begin{eqnarray}
g(\mathbf{p},s)= && \!\!\!
\frac{\delta_{ss^0}}{(\sqrt{\frac{\pi}{2}}\sigma)^{3/2}}
\mathrm{e}^{-(\mathbf{p}-\mathbf{p}^0)^2/\sigma^2} \nonumber,
\end{eqnarray}
which in the limit of vanishing widths give the standard -well
defined- momentum state $\delta_{s
s^0}\delta^{(3)}(\mathbf{p}-\mathbf{p}^0)$.

In the absence of interactions, a separable initial state will
continue to be separable forever. However, interactions destroy
this simple picture due to the effect of U. Clearly, the final
state
\begin{eqnarray}
F^{(2)}(\mathbf{p}_1,s_1;\mathbf{p}_2,s_2;t) \!\!\! & = & \!\!\!\!
\sum_{s_a,s_b} \!\! \int \!\! d^3\mathbf{p}_a \! \int \!\!
d^3\mathbf{p}_b
\tilde{K}^{(2)}(1,2;a,b;t) \nonumber \\
&\times& F(\mathbf{p}_a,s_a;\mathbf{p}_b,s_b)\label{teeglo1}
\end{eqnarray}
can not be factorized.

In the rest of this section we analyze the final state
$F^{(2)}(\mathbf{p}_1,s_1;\mathbf{p}_2,s_2;t)$ in
Eq.~(\ref{teeglo1}) to show how the variables $\mathbf{p}_1$ and
$\mathbf{p}_2$ get entangled by the interaction. We consider the
nonrelativistic regime in which all intervening momenta and widths
$\mathbf{p},\sigma\ll m$, so the characteristic times $t$ under
consideration are appreciable. We single out the particular case
of a projectile fermion $a$ scattered off a fuzzy target fermion
$b$ centered around $\mathbf{p}_b^0=0$. As a further
simplification, we consider the projectile momentum sharply
distributed around $\mathbf{p}_a^0$ ($\sigma_a\ll\mathbf{p}_a^0$)
so that the initial state can be approximated by
\begin{equation}
F(a;b)\approx
\delta_{s_as_a^0}\delta^{(3)}(\mathbf{p}_a-\mathbf{p}^0_a)
\frac{\delta_{s_bs_b^0}}{(\sqrt{\frac{\pi}{2}}\sigma_b)^{3/2}}
\mathrm{e}^{-(\mathbf{p}_b-\mathbf{p}^0_b)^2/\sigma_b^2} .
\label{eqges4}
\end{equation}
Our kinematical configuration would acquire complete generality
should we introduce a finite momenta $\mathbf{p}^0_b$ for the
initial electron b. The reference system would be in this case
midway between the lab. system and the c.o.m. system. In short,  the
choice $\mathbf{p}^0_b = 0$ will not affect the qualitative
properties of entanglement generation.

\begin{figure}
\begin{center}
\includegraphics[width=0.4\textwidth]{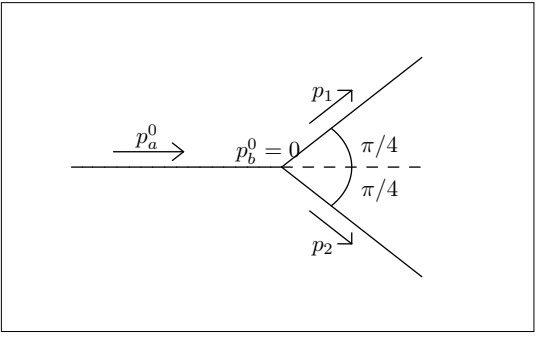}
\end{center}
\caption{Experimental setup considered in the calculations.
\label{figgesinicial}}
\end{figure}

We will work in the lab frame, where particle $b$ shows a fuzzy
momentum distribution around $\mathbf{p}^0_b=0$, and focus in the
kinematical situation in which the final state momenta satisfy
$\mathbf{p}_1\cdot\mathbf{p}_2=0$ and also
$\mathbf{p}_{\alpha}\cdot\mathbf{p}_a^0=1/\sqrt{2}p_{\alpha}p_a^0$,
$\alpha=1,2$ (see Fig. \ref{figgesinicial}). This choice not only
avoids forward scattering divergencies but also simplifies the
expression of the amplitude in Eq.~(\ref{teeglo1}), due to the
chosen angles. For sure, the qualitative conclusions would also
hold in other frames, like the center-of-mass one.
We obtain
\begin{eqnarray}
&& \!\!\!\!\!\!
E_1+E_2-E_a-E_b|_{\mathbf{p}_a=\mathbf{p}_a^0}^{\mathbf{p}_b=\mathbf{p}_1
+\mathbf{p}_2-\mathbf{p}_a^0} = \nonumber \\ &&
\frac{p_a^0}{\sqrt{2}m}(p_1+p_2-\sqrt{2}p_a^0)+O((p_a^0/m)^3p_a^0)
, \nonumber \\ &&
\frac{(\mathbf{p}_1+\mathbf{p}_2-\mathbf{p}_a^0)^2}{\sigma^2} =
\frac{(p_1-p_a^0/\sqrt{2})^2}{\sigma^2}
+\frac{(p_2-p_a^0/\sqrt{2})^2}{\sigma^2} , \nonumber \\ &&
(\mathbf{p}_1-\mathbf{p}_a)^2 = (p_1-p_a^0/\sqrt{2})^2+(p_a^0)^2/2
, \nonumber \\ && (\mathbf{p}_2-\mathbf{p}_a)^2 =
(p_2-p_a^0/\sqrt{2})^2+(p_a^0)^2/2. \label{eqges9}
\end{eqnarray}
Here, boldface characters represent trivectors, otherwise they
represent their associated norms. We perform now the following
change of variables,
\begin{equation}
\frac{p}{\sqrt{2}}=\frac{1}{\sigma}\left(p_1-\frac{p_a^0}{\sqrt{2}}
\right) , \;\;\;\;
\frac{q}{\sqrt{2}}=\frac{1}{\sigma}\left(p_2-\frac{p_a^0}{\sqrt{2}}
\right) ,\label{eqges10}
\end{equation}
turning the amplitude in Eq.~(\ref{teeglo1}) into
\begin{eqnarray}
 && F^{(2)}(p,s_1;q,s_2;t) \propto\frac{\sin[(p+q)\tilde{t}]}
{\tilde{\Sigma}}\nonumber\\&\times&\left[\frac{(j^{\mu}_{1a}j_{\mu
2b})^{s_a=s_a^0}_{s_b=s_b^0}}{p^2+\left(\frac{p_a^0}{\sigma}
\right)^2}-\frac{(j^{\mu}_{1b}j_{\mu
2a})^{s_a=s_a^0}_{s_b=s_b^0}}{q^2+\left(\frac{p_a^0}{\sigma}
\right)^2}\right]\mathrm{e}^{-p^2/2}\mathrm{e}^{-q^2/2}\nonumber\\
&+&\left(\frac{(j^{\mu}_{1a}j_{\mu
2b})^{s_a=s_a^0}_{s_b=s_b^0}}{\tilde{\mu}/2}\right.
\left\{-\frac{1}{\tilde{\mu}(\tilde{\Sigma}^2-\tilde{\mu}^2)}
\right.\tilde{\Sigma}\sin[(p+q)\tilde{t}]\nonumber\\&-&\left.
\frac{i}{\tilde{\Sigma}^2-\tilde{\mu}^2}
\left(\cos[(p+q)\tilde{t}]-\mathrm{e}^{-i\frac{2m}{p_a^0}
\tilde{\mu}\tilde{t}}\right)\right\}
\nonumber\\&-&\Biggl.\{p,1\leftrightarrow
q,2\}\Biggr)\mathrm{e}^{-p^2/2}\mathrm{e}^{-q^2/2},
\label{eqges11}
\end{eqnarray}
where $\tilde{\Sigma}=\frac{p_a^0}{2m}(p+q)$,
$\tilde{\mu}=\sqrt{2}\sqrt{p^2+\left(\frac{p_a^0}{\sigma}\right)^2}$,
and $\tilde{t}\equiv\frac{p_a^0\sigma}{2m}t$. In the following, we
analyze different specific spin configurations in the
non-relativistic limit with the help of Eq.~(\ref{eqges11}). We
consider an incident particle energy of around $1$ eV$\ll m$
($p_a^0=1$ KeV), and a momentum spreading $\sigma$ one order of
magnitude less than $p_a^0$. We make this choice of $p_a^0$ and
$\sigma$ to obtain longer interaction times, of femtoseconds
($t=\frac{2m}{p_a^0\sigma}\tilde{t}$). Thus the parameter values
we consider in the subsequent analysis are $p_a^0/m=0.002$ and
$\sigma/m=0.0002$. We consider the initial spin state for
particles $a$ and $b$ as
\begin{eqnarray}
|s_a^0s_b^0\rangle =|\uparrow\downarrow\rangle ,
\label{eqsges1}
\end{eqnarray}
along an arbitrary direction that will serve to measure spin
components in all the calculation. The physical results we are
interested in do not depend on this choice of direction. The QED
interaction, in the non-relativistic regime considered, at lowest
order, is a Coulomb interaction that does not change the spins of
the fermions. In fact, $(j^{\mu}_{1a}j_{\mu 2b})\simeq
4m^2\delta_{s_a^0s_1}\delta_{s_b^0s_2}$, $(j^{\mu}_{1b}j_{\mu
2a})\simeq 4m^2\delta_{s_b^0s_1}\delta_{s_a^0s_2}$. Given the
initial spin states of Eq.~(\ref{eqsges1}), depending on whether
the channel is $t$ or $u$, the possible final spin states are
\begin{eqnarray}
|s_1s_2\rangle_t &=&
|\uparrow\downarrow\rangle,\label{eqsges1bis}\\
|s_1s_2\rangle_u&=&|\downarrow\uparrow\rangle .
\label{eqsges1bisbis}
\end{eqnarray}
Due to the fact that the considered fermions are identical, the
resulting amplitude after applying the Schmidt procedure is a
superposition of Slater determinants
\cite{entanglefermion1,ESB02,entanglefermion2}. Whenever this
decomposition contains just one Slater determinant (Slater number
equal to 1) the state is not entangled: its correlations are just
due to the statistics and are not useful for the applications
because they do not contain any additional physical information.
If the amplitude contains more than one determinant, the state is
entangled. Splitting the amplitude in the corresponding ones for
the $t$ and $u$ channels, we have


\begin{eqnarray}
&&F^{(2)}(p,\uparrow;q,\downarrow;t)_t\propto \nonumber \\ &&
\frac{\sin[(p+q)\tilde{t}]}{\tilde{\Sigma}}\frac{1}{p^2
+\left(\frac{p_a^0}{\sigma}\right)^2}
\mathrm{e}^{-p^2/2}\mathrm{e}^{-q^2/2}\nonumber\\
&+&\frac{1}{\tilde{\mu}/2}\left\{-\frac{1}{\tilde{\mu}
(\tilde{\Sigma}^2-\tilde{\mu}^2)}
\right.\tilde{\Sigma}\sin[(p+q)\tilde{t}]\nonumber
\\ & - & \left. \frac{i}{\tilde{\Sigma}^2-\tilde{\mu}^2}
\left(\cos[(p+q)\tilde{t}]-\mathrm{e}^{-i\frac{2m}{p_a^0}
\tilde{\mu}\tilde{t}}\right)\right\} \nonumber
\\ & \times & \mathrm{e}^{-p^2/2}\mathrm{e}^{-q^2/2},
\label{eqsges3}
\end{eqnarray}
with
\begin{eqnarray}
& F^{(2)}(p,\downarrow;q,\uparrow;t)_u \leftrightarrow
F^{(2)}(p,\uparrow;q,\downarrow;t)_t , & \nonumber
\\ & p \leftrightarrow q . & \label{eqsges3bis}
\end{eqnarray}

In the infinite time limit the sinc function converges to
$\delta(p+q)$, which is a distribution with infinite entanglement
\cite{lljl}. The presence of the sinc function is due to the finite
time interval of integration in Eq.~(\ref{eqfte12}). This kind of
behavior can be interpreted as a time diffraction
phenomenon~\cite{moshinsky}. It has direct analogy with the
diffraction of electromagnetic waves that go through a single slit
of width $2L$ comparable to the wavelength $\lambda$. The analogy is
complete if one identifies $\tilde{t}$ with $L$ and $p+q$ with
$2\pi/\lambda$.

\begin{figure}[h]
\begin{center}
\includegraphics{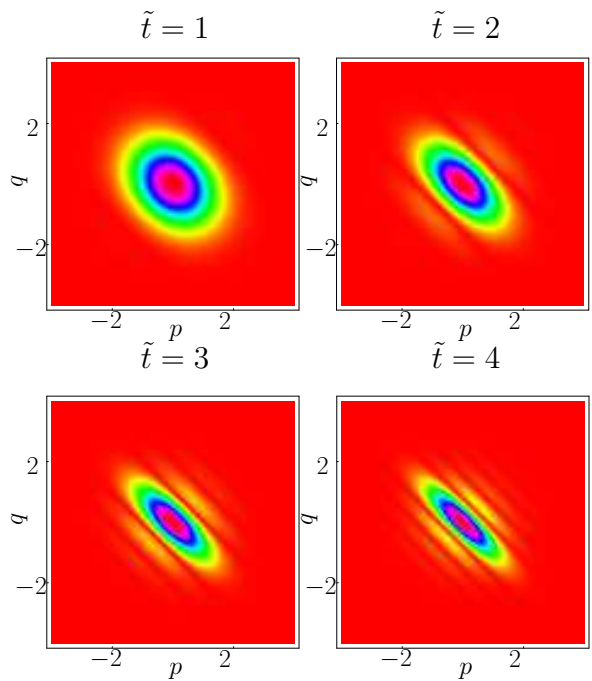}
\end{center}
\caption{(Color online) $|F^{(2)}(p,\uparrow;q,\downarrow;t)_t|$
versus $p$, $q$ at $\tilde{t}=1,2,3,4$ . \label{figsqed1}}
\end{figure}

In Fig.~\ref{figsqed1}, we plot the modulus of Eq.~(\ref{eqsges3})
versus $p$, $q$, at times $\tilde{t}=1,2,3,4$. This graphic shows
the progressive clustering of the amplitude around the curve
$q=-p$, due to the function $\frac{\sin[(p+q)\tilde{t}]}{p+q}$.
This is a clear signal of the growth in time of the momentum
entanglement. Fig.~\ref{figsqed1} puts also in evidence the
previously mentioned time diffraction pattern.

\begin{figure}
\begin{center}
\includegraphics{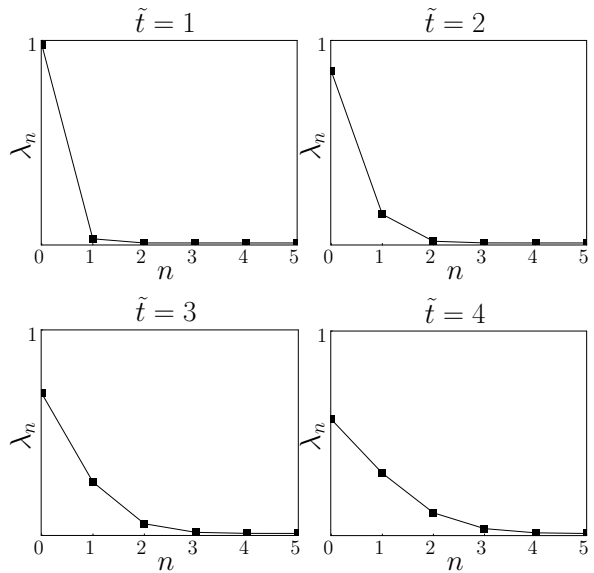}
\end{center}
\caption{Eigenvalues $\lambda_n$ versus $n$ at times $\tilde{t}
=1,2,3,4$.\label{figsqed2}}
\end{figure}

We have applied the method for obtaining the Schmidt decomposition
given in Ref.~\cite{lljl} to Eq.~(\ref{eqsges3}), considering for
the orthonormal functions $\{O^{(1)}(p)\}$, $\{O^{(2)}(q)\}$
Hermite polynomials with their weights, to take advantage of the
two Gaussian functions. We obtain the Schmidt decomposition for
$\tilde{t}=1,2,3,4$, where the error with matrices $C_{mn}$
$12\times 12$ or smaller is $d^{2}_{m_0,n_0}\leq 7 \cdot10^{-3}$
in all considered cases. We plot in Fig.~\ref{figsqed2} the
coefficients $\lambda_n$ of the Schmidt decomposition of
Eq.~(\ref{eqsges3}) as a function of $n$, for times
$\tilde{t}=1,2,3,4$. The number of $\lambda_n$ different from zero
increases as time is elapsed, and thus the entanglement grows.

The complete Schmidt decomposition, including channels $t$ and
$u$, is given in terms of Slater determinants
\cite{entanglefermion1}, and is usually called Slater
decomposition. It is obtained antisymmetrizing the amplitude for
channel $t$
\begin{eqnarray}
&&F^{(2)}(p,s_1;q,s_2;t) \propto \sum_n
\sqrt{\lambda_n(\tilde{t})}\nonumber\\&\times&\frac{\psi^{(1)}_n
(p,\tilde{t})|\uparrow\rangle\psi^{(2)}_n(q,\tilde{t})|\downarrow
\rangle
-\psi^{(2)}_n(p,\tilde{t})|\downarrow\rangle\psi^{(1)}_n(q,\tilde{t})
|\uparrow\rangle}{\sqrt{2}},\nonumber\\\label{eqges13bis}
\end{eqnarray}
where the modes $\psi^{(1)}_n(k,\tilde{t})$ and
$\psi^{(2)}_n(k,\tilde{t})$ are the Schmidt modes of the channel
$t$ obtained for particles $1$ and $2$ respectively, and they
correspond to the modes of the channel $u$ for particles $2$ and
$1$ respectively.

\begin{figure}
\begin{center}
\includegraphics[width=6cm]{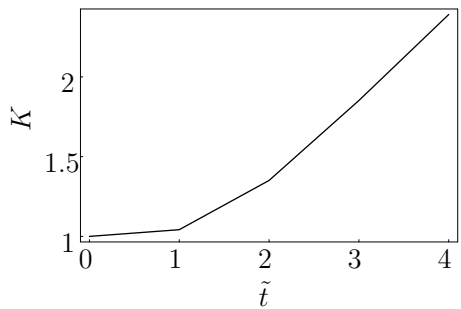}
\end{center}
\caption{Slater number $K$ as a function of the elapsed time
$\tilde{t}$.\label{figsqed3}}
\end{figure}

A measure of the entanglement of a pure bipartite state of the form
of Eq.~(\ref{eqges13bis}), equivalent to the entropy of entanglement
$S$, is given by the Slater number \cite{qedentang}
\begin{equation}
K\equiv \frac{1}{\sum_{n=0}^{\infty}\lambda_n^2} \,
.
\label{eqsges4}
\end{equation}
$K$ gives the number of effective Slater determinants which appear
in a certain pure bipartite state in the form of
Eq.~(\ref{eqges13bis}). The larger the value of $K$, the larger the
entanglement. For $K=1$ (one Slater determinant) there is no
entanglement. This measure is obtained considering the average
probability, which is given by $\sum_{n=0}^{\infty}\lambda_n^2$
($\sum_{n=0}^{\infty}\lambda_n=1$, and thus $\{\lambda_n\}$ can be
seen as a probability distribution). The inverse of the average
probability is the Slater number. Its attractive properties are that
it is independent of the representation of the wavefunction, it is
gauge invariant, and it reaches its minimum value of 1 for the
separable state (single Slater determinant). In Fig. \ref{figsqed3},
we show the Slater number $K$ as a function of elapsed time
$\tilde{t}$, verifying that the entanglement increases as the system
evolves. It can be appreciated in this figure the monotonic growth
of entanglement, due to the fact that we have considered an incident
electron with well defined momentum. In realistic physical
situations with wave packets, this growth would stop, due to the
momentum spread of the initial electrons.  The general trend is that
the higher the precision in the incident electron momentum, the
larger the resulting asymptotic entanglement. The fact that the
entanglement in momenta between the two fermions increases with time
is a consequence of the interaction between them. We remark that the
entanglement cannot grow unless the two particles ``feel'' each
other. The correlations in momenta are not specific of QED: the
effect of any interaction producing momentum exchange while
conserving total momentum will translate into momentum correlations.

The Schmidt modes in momenta space for the amplitude of
Eq.~(\ref{eqsges3}) are given by
\begin{eqnarray}
\psi^{(\alpha)}_m(k,\tilde{t})&\simeq&\mathrm{e}^{-k^2/2}
\sum_{n=0}^{n_0}(\sqrt{\pi}2^nn!)^{-1/2}
A^{(\alpha)}_{mn}(\tilde{t})H_n(k)\;\;\;\; \nonumber \\ \alpha
&=&1,2, \label{eqges13}
\end{eqnarray}
where $n_0$ is the corresponding cut-off and the values of the
coefficients $A^{(\alpha)}_{mn}(\tilde{t})$ are obtained through
the method given in Ref.~\cite{lljl}. The modes in momenta space
depend on time because they are not stationary states: the QED
dynamics between the two fermions and the indeterminacy on the
energy at early stages of the interaction give this dependence. By
construction, the coefficients $A^{(\alpha)}_{mn}(\tilde{t})$ do
not depend on $p$, $q$.

\begin{figure}
\begin{center}
\includegraphics{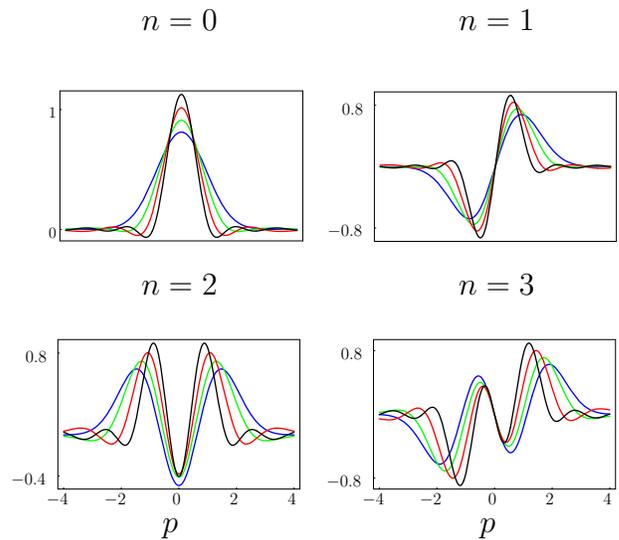}
\end{center}
 \caption{(Color online) Schmidt modes $\psi^{(1)}_n(p,\tilde{t})$ at times
 $\tilde{t}=1,2,3,4$ for $n=0,1,2,3$. The sharper modes for each $n$
 correspond to the later
times.\label{figsqed4}}
\end{figure}

We plot in Fig.~\ref{figsqed4} the Schmidt modes
$\psi^{(1)}_n(p,\tilde{t})$ at times $\tilde{t}=1,2,3,4$ for
$n=0,1,2,3$ (we are plotting specifically the real part of each
mode only, which approximates well the whole mode, because
Eq.~(\ref{eqsges3}) is almost real for the cases considered). The
sharper modes for each $n$ correspond to the later times. Each
Schmidt mode is well approximated at early times by the
corresponding Hermite orthonormal function, and afterwards it
sharpens and deviates from that function: it gets corrections from
higher order polynomials. The fact that the modes get thinner with
time is related to the behavior of Eq.~(\ref{eqsges3}) at large
times. In particular the sinc function goes to $\delta(p+q)$ and
thus the amplitude gets sharper.

Now we consider the Schmidt modes in configuration space. To
obtain them, we just Fourier transform the modes of
Eq.~(\ref{eqges13}) with respect to the momenta $p_1$, $p_2$
\begin{equation}
\tilde{\psi}^{(\alpha)}_m(x_{\alpha},\tilde{t})=\frac{1}{\sqrt{2\pi}}
\int_{-\infty}^{\infty}dp_{\alpha}\mathrm{e}^{i(p_{\alpha}x_{\alpha}
-\frac{p_{\alpha}^2}{2m}t)}
\psi^{(\alpha)}_m(k(p_{\alpha}),\tilde{t}),\label{eqges14}
\end{equation}
where $\alpha=1,2$. The dependence of $p$ on $p_1$ and $q$ on
$p_2$ is given through Eq.~(\ref{eqges10}). The factor
$\mathrm{e}^{-i\frac{p_{\alpha}^2}{2m}t}$ in Eq.~(\ref{eqges14})
is the one which commutes the states between the interaction
picture (considered in Eq.~(\ref{eqges13}) and in the previous
calculations in Secs. \ref{fte} and \ref{ges}) and the
Schr\"{o}dinger picture.

The Hermite polynomials obey the following expression \cite{grads}
\begin{equation}
\int_{-\infty}^{\infty}dx \mathrm{e}^{-(x-y)^2}H_n(\alpha
x)=\sqrt{\pi}(1-\alpha^2)^{n/2}H_n\left(\frac{\alpha
y}{\sqrt{1-\alpha^2}}\right).\label{eqges15}
\end{equation}
With the help of Eq.~(\ref{eqges15}) and by linearity of the
Fourier transforms, we are able to obtain analytic expressions for
the Schmidt modes in configuration space (to a certain accuracy,
which depends on the cut-offs considered). This is possible
because the dispersion relation of the massive fermions in the
considered non-relativistic limit is
$E_{\alpha}=\frac{p_{\alpha}^2}{2m}$, and thus the integral of
Eq.~(\ref{eqges14}) can be obtained analytically using
Eq.~(\ref{eqges15}).

The corresponding Schmidt modes in configuration space are then
given by
\begin{eqnarray}
\tilde{\psi}^{(\alpha)}_m(\tilde{x}_{\alpha},\tilde{t})\simeq
\sum_{n=0}^{n_0}A^{(\alpha)}_{mn}(\tilde{t})
\tilde{O}^{(\alpha)}_n(\tilde{x}_{\alpha},\tilde{t}) , \;\;\;\;
\alpha=1,2, \label{eqges16}
\end{eqnarray}
where the orthonormal functions in configuration space are
\begin{eqnarray}
\tilde{O}^{(\alpha)}_n(\tilde{x}_{\alpha},\tilde{t}) \!\! &
\!\!\!\! = \!\!\!\!
 & \!\! i^n (\sqrt{\pi}2^nn!)^{-1/2}
\frac{\mathrm{e}^{-in\arctan(\tilde{\sigma}\tilde{t})
+i\tilde{\sigma}^{-1}(\tilde{x}_{\alpha}-\tilde{t}/2)}}
{\sqrt{1+i\tilde{\sigma}\tilde{t}}}\nonumber\\&\times&
\mathrm{e}^{-\frac{(\tilde{x}_{\alpha}-\tilde{t})^2}
{2(1+i\tilde{\sigma}\tilde{t})}}
H_n\left[\frac{\tilde{t}-\tilde{x}_{\alpha}}{\sqrt{1
+(\tilde{\sigma}\tilde{t})^2}}\right].\label{eqges17}
\end{eqnarray}
In Eqs.~(\ref{eqges16}) and (\ref{eqges17}), we are using
dimensionless variables, $\tilde{x}_{\alpha}=\frac{\sigma
x_{\alpha}}{\sqrt{2}}$, $\alpha=1,2$,
$\tilde{\sigma}=\frac{\sigma}{p_a^0}$, and the dimensionless time
defined before, $\tilde{t}=\frac{p_a^0\sigma}{2m}t$. The modes in
Eqs.~(\ref{eqges16}) and (\ref{eqges17}) are normalized in the
variables $\tilde{x}_{\alpha}$. The orthonormal functions of
Eq.~(\ref{eqges17}) propagate in space at a speed
$\frac{p_a^0}{\sqrt{2}m}$ and they spread in their evolution.
Additionally, the modes of Eq.~(\ref{eqges16}) have also the time
dependence of $A^{(\alpha)}_{mn}(\tilde{t})$. The Slater
decomposition in configuration space, obtained Fourier
transforming the modes of Eq.~(\ref{eqges13}) is
\begin{eqnarray}
\tilde{F}^{(2)}(\tilde{x}_1,\tilde{x}_2,\tilde{t})&\propto&\sum_n
\sqrt{\frac{\lambda_n(\tilde{t})}{2}}[\tilde{\psi}^{(1)}_n
(\tilde{x}_1,\tilde{t})|\uparrow\rangle\tilde{\psi}^{(2)}_n
(\tilde{x}_2,\tilde{t})|\downarrow\rangle
\nonumber\\&-&\tilde{\psi}^{(2)}_n(\tilde{x}_1,\tilde{t})|
\downarrow\rangle\tilde{\psi}^{(1)}_n(\tilde{x}_2,\tilde{t})
|\uparrow\rangle].\label{eqges18}
\end{eqnarray}
The coefficients $\lambda_n(\tilde{t})$ are unaffected by the
Fourier transformation, and thus the degree of entanglement in
configuration space is the same as in momenta space.

We consider now the initial spin configuration
\begin{eqnarray}
|s_a^0s_b^0\rangle=|\uparrow\uparrow\rangle ,
\label{eqtges1}
\end{eqnarray}
where, the only possible final state in the non-relativistic limit
is
\begin{eqnarray}
|s_1s_2\rangle=|\uparrow\uparrow\rangle.\label{eqtges2}
\end{eqnarray}
In this case, the sinc term goes to zero, because the momentum part
of this term is antisymmetric in $p^2$, $q^2$ and the sinc function
goes to $\delta(p+q)$, which has support (as a distribution) on
$q=-p$. We point out that the sinc contribution to this amplitude is
negligible because of the particular setup chosen. In other
experiment configurations the amplitude in Eq.~(\ref{teeglo1})
associated to the spin states of Eqs.~(\ref{eqtges1}) and
(\ref{eqtges2}) would have appreciable sinc term and thus increasing
momenta entanglement with time. On the other hand, in this case the
contribution from $\Upsilon_t(t)$ in Eq.~(\ref{eqfte17bis}) and
$\Upsilon_u(t)$ is even smaller than the sinc term, and converges
weakly to zero.

\begin{figure}[h]
\begin{center}
 \includegraphics [width=7cm]{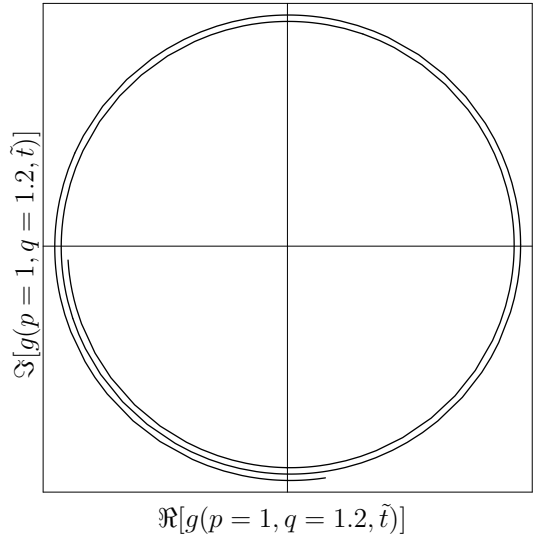}
\end{center}
 \caption{Real and imaginary parts of the amplitude $g(p,q,
 \tilde{t})$ for $p=1$, $q=1.2$, and
$\tilde{t}\in(1,1.001)$ (arbitrary units).\label{figtqed2}}
\end{figure}

We plot in Fig.~\ref{figtqed2} the real and imaginary parts of the
term associated to $\Upsilon_t(t)$ and $\Upsilon_u(t)$ in
Eq.~(\ref{eqges11}), which we denote by $g(p,q,\tilde{t})$, for
spin states of Eqs.~(\ref{eqtges1}) and (\ref{eqtges2}) as a
function of time $\tilde{t}\in(1,1.001)$ and having $p=1$,
$q=1.2$. We want to show with it the strong oscillatory character
of the amplitude with time, and how all the contributions
interfere destructively with each other giving a zero final value.
This is similar to the stationary phase procedure, in which only
the contributions in proximity to the stationary value of the
phase do interfere constructively and are appreciable. What we
display here is the weak convergence to zero for the functions
$\Upsilon_t(t)$ and $\Upsilon_u(t)$.

In this section, we investigated the generation of entanglement in
momenta between two identical spin-1/2 particles which interact via
QED. We showed how the correlations grow as the energy conservation
is increasingly fulfilled with time. The previous calculation had,
however, the approximation of considering a projectile particle with
perfectly well defined momentum, something not achievable in
practice. This is a first step towards a real experiment, where both
fermions will have a dispersion in momenta and thus infinite
entanglement will never be reached, due to the additional integrals
of the Dirac delta $\delta(\Delta E)$ over the momentum spread.

\section{Conclusions}

We analyzed the dynamical generation of  entanglement between two
electrons due to their mutual interaction by means of the Slater
number. In the asymptotic limit, and for electrons whose initial
momenta are sharply defined, entanglement divergencies may appear.
We observe that considering finite-time intervals of interaction,
and/or a certain spreading in the particles momentum, the
entanglement remains finite. We have studied for the first time
the dynamical generation of momentum entanglement of two spin-1/2
particles at lowest order QED, observing how the correlations
increase as the particles exchange virtual photons. We obtain the
Schmidt decomposition of the scattering amplitude, given in terms
of Slater determinants, and the Slater number, which clearly shows
the growth of entanglement with time. We also obtain analytic
approximations of the Schmidt modes, both in momentum and
configuration spaces.

\section*{ACKNOWLEDGEMENTS}

J.L. thanks I. Bialynicki-Birula for useful discussions and private
communications. L.L. thanks I. Cirac for hospitality at Max-Planck
Institute for Quantum Optics and acknowledges financial support from
FPU grant AP2003-0014. The work of J.L. and L.L. was partially
supported by the Spanish Ministerio de Educaci\'on y Ciencia under
project BMF 2002-00834. E.S. acknowledges financial support from EU
Project RESQ.

\appendix

\section{Entanglement in finite and infinite dimensional Hilbert
spaces \label{entangfi}}

We consider a composite system ${\mathcal S}$ described by a
Hilbert space $\mathcal{H}$, which may be either finite or
infinite dimensional. This space is constructed as the tensor
product of the Hilbert spaces associated to each of the
subsystems, ${\mathcal S}_{\alpha}$, of ${\mathcal S}$. For
simplicity, and in view of our present purposes, we restrict
ourselves to pure states of a bipartite system ${\mathcal S}$.
Thus, $\alpha=1,2$, and
$\mathcal{H}=\mathcal{H}_1\otimes\mathcal{H}_2$.

{\it Definition: product state}. A vector state $|\Psi\rangle$ of
the system ${\mathcal S}$ is a product state if it can be written
as
\begin{equation}
|\Psi\rangle=|\Psi^{(1)}\rangle |\Psi^{(2)}\rangle ,
\label{eqentan1}
\end{equation}
where $|\Psi^{(1)}\rangle \in \mathcal{H}_1$ and
$|\Psi^{(2)}\rangle \in \mathcal{H}_2$.

{\it Definition: entangled state}. A vector state $|\Psi\rangle$
of the system ${\mathcal S}$ is entangled if it is not a product
state. A relevant example, where
$\dim(\mathcal{H}_1)=\dim(\mathcal{H}_2)=2$, is called the singlet
state,
\begin{equation}
|\Psi^-\rangle=\frac{1}{\sqrt{2}}(|\Psi^{(1)}_1\rangle
|\Psi^{(2)}_2\rangle-|\Psi^{(1)}_2\rangle
|\Psi^{(2)}_1\rangle).\label{eqentan2}
\end{equation}
A very useful tool for analyzing the entanglement of pure states
of bipartite systems is given by the Schmidt decomposition
\cite{schmidtdisc2, eberly1}. It basically consists in expressing
the pure bipartite state as a sum of biorthonormal products, with
positive coefficients $\sqrt{\lambda_n}$, as follows
\begin{equation}
|\Psi\rangle=\sum_{n=0}^{d-1} \sqrt{\lambda_n}|\Psi^{(1)}_n\rangle
|\Psi^{(2)}_n\rangle\label{eqentan3},
\end{equation}
where $\{|\Psi^{(1)}_n\rangle\}$, $\{|\Psi^{(2)}_n\rangle\}$, are
orthonormal bases associated to $\mathcal{H}_1$ and
$\mathcal{H}_2$, respectively. In Eq.~(\ref{eqentan3}), $d= \min
\{ \dim(\mathcal{H}_1),\dim(\mathcal{H}_2) \}$, and it may be
infinite, as for in continuous variable systems, describing
momentum, energy, position, frequency, or the like. In those
cases, the states $|\Psi^{(\alpha)}_n\rangle$ would be (square
integrable) $L^2$ wave functions,
\begin{eqnarray}
\langle p|\Psi^{(\alpha)}_n\rangle \! & = & \!
\psi^{(\alpha)}_n(p) , \;\;\; \alpha = 1,2, \label{eqentan4}
\end{eqnarray}
where $p$ denotes the corresponding continuous variable.

For pure bipartite states a relevant measure of entanglement is
the {\it entropy of entanglement}, $S$. Given a state
$|\Psi\rangle$, it is defined as the von Neumann entropy of the
reduced density matrix with respect to $S_1$ or $S_2$,
\begin{equation}
S=-\sum_{n=0}^{d-1}\lambda_n\log_2\lambda_n,\label{eqentan5}
\end{equation}
where the $\lambda_n$'s are given in Eq.~(\ref{eqentan3}). In
general, $S\geq0$, $S = 0$ for a product state, and the more
entangled a state is, the larger $S$. For a maximally entangled
state, $S=\log_2 d$, and if $d=\infty$, then $S$ diverges.

An interesting work where the Schmidt decomposition for continuous
variables is analyzed discretizing the corresponding integral
equations can be found in Ref.~\cite{eberly1}. Another method for
obtaining the continuous variables Schmidt decomposition, based in
decomposing the bipartite wave function in complete sets of
orthonormal functions, is developed in Ref.~\cite{lljl}.

\section{Entanglement transfer between momentum and spin\label{maj}}

\subsection{Dynamical transfer and distillation }

In Sec.\ref{ges} we computed the entanglement in momenta for a
pair of identical spin-1/2 particles which interact through
exchange of a virtual photon. The sharper the initial momentum
distribution of the incident fermion, and the longer the
interaction time, the larger the entanglement in momenta.
Heisenberg's principle, on the other hand, establishes a limit to
the precision with which the momentum may be defined and hence to
the achievable degree of entanglement.

It is possible, in principle, to transform the entanglement in
momenta into entanglement in spins. This is easily seen in terms of
the majorization criterion~\cite{majoriz,NielsenChuang}, which is of
practical interest because the experimentalist usually manipulates
spins. Here, we will analyze this entanglement transfer.

Majorization is an area of mathematics which predates quantum
mechanics. Quoting Nielsen and Chuang, ``Majorization is an ordering
on d-dimensional real vectors intended to capture the notion that
one vector is more or less disordered than another''. We consider a
pair of $d$-dimensional vectors, $x=(x_1,...,x_d)$ and
$y=(y_1,...,y_d)$. We say $x$ is majorized by $y$, written $x\prec
y$, if $\sum_{j=1}^k x_j^{\downarrow}\leq\sum_{j=1}^k
y_j^{\downarrow}$ for $k=1,...,d$, with equality instead of
inequality for $k=d$. We denote by $z^{\downarrow}$ the components
of $z$ in decreasing order $(z_1^{\downarrow}\geq
z_2^{\downarrow}\geq...\geq z_d^{\downarrow})$. The interest of this
work in the majorization concept comes from a theorem which states
that a bipartite pure state $|\psi\rangle$ may be transformed to
another pure state $|\phi\rangle$ by Local Operations and Classical
Communication (LOCC) if and only if
$\lambda_{\psi}\prec\lambda_{\phi}$, where $\lambda_{\psi}$,
$\lambda_{\phi}$ are the vectors of (square) coefficients of the
Schmidt decomposition of the states $|\psi\rangle$, $|\phi\rangle$,
respectively. LOCC adds to those quantum operations effected only
locally the possibility of classical communication between spatially
separated parts of the system. According to this criterion, it would
be possible in principle to obtain a singlet spin state
$|\phi\rangle$ beginning with a momentum entangled state
$|\psi\rangle$ whenever $\lambda_{\psi}\prec\lambda_{\phi}$.

The possibility of obtaining a singlet spin state from a
momentum-entangled state can be extended to a more efficient
situation: the possibility of distillation of entanglement. This
idea consists on obtaining multiple singlet states beginning with
several copies of a given pure state $|\psi\rangle$. The distillable
entanglement of $|\psi\rangle$ consists in the ratio $n/m$, where
$m$ is the number of copies of $|\psi\rangle$ we have initially, and
$n$ the number of singlet states we are able to obtain via LOCC
acting on these copies. It can be shown \cite{NielsenChuang} that
for pure states the distillable entanglement equals the entropy of
entanglement, S. Thus, in the continuous case (infinite-dimensional
Hilbert space), the distillable entanglement is not bounded from
above, because neither is S. According to this, the larger the
entanglement in momenta the more singlet states could be obtained
with LOCC.

To illustrate the possibility of entanglement transfer with a
specific example, we consider a momentum-entangled state for two
distinguishable fermions
\begin{equation}
|\psi\rangle=\frac{1}{\sqrt{2}}[\psi^{(1)}_1(p)\psi^{(2)}_1(q)
+\psi^{(1)}_2(p)\psi^{(2)}_2(q)]
|\uparrow\uparrow\rangle.\label{eqmced1}
\end{equation}
This state has associated a vector
$\lambda_{\psi}^{\downarrow}=(1/2,1/2,0,0,...)$. On the other hand,
the singlet state
\begin{equation}
|\phi\rangle=\psi^{(1)}_1(p)\psi^{(2)}_1(q)
\frac{1}{\sqrt{2}}(|\uparrow\downarrow\rangle-|\downarrow
\uparrow\rangle)\label{eqmced2}
\end{equation}
has associated a vector
$\lambda_{\phi}^{\downarrow}=(1/2,1/2,0,0...)$.

These vectors obey $\lambda_{\psi}\prec\lambda_{\phi}$, and thus the
state entangled in momenta may be transformed into the state
entangled in spins via LOCC.

\subsection{Kinematical transfer and Lorentz boosts}

Another approach to the study of entanglement transfer between
momentum and spin degrees of freedom is the kinematical one. In
fact, the Lorentz transformations may entangle the spin and momentum
degrees of freedom. To be more explicit, and following the notation
of Ref.~\cite{GA02}, we consider a certain bipartite pure wave
function $g_{\lambda\sigma}(\mathbf{p},\mathbf{q})$ for two spin-1/2
fermions, where $\lambda$ and $\sigma$ denote respectively the spin
degrees of freedom of each of the two fermions, and $\mathbf{p}$ and
$\mathbf{q}$ the corresponding momenta. This would appear to an
observer in a Lorentz transformed frame as
\begin{equation}
g_{\lambda\sigma}(\mathbf{p},\mathbf{q})\begin{array}{c}\Lambda \\
\longrightarrow\end{array}
\sum_{\lambda'\sigma'}U_{\lambda\lambda'}^{(\Lambda^{-1}\mathbf{p})}
U_{\sigma\sigma'}^{(\Lambda^{-1}\mathbf{q})}g_{\lambda'\sigma'}
(\Lambda^{-1}\mathbf{p},\Lambda^{-1}\mathbf{q}) , \label{eqmced4}
\end{equation}
where
\begin{equation}
U_{\lambda\lambda'}^{(\mathbf{p})}\equiv
D_{\lambda\lambda'}^{(1/2)}(R(\Lambda,\mathbf{p}))\label{eqmced5}
\end{equation}
is the spin 1/2 representation of the Wigner rotation
$R(\Lambda,\mathbf{p})$. The Wigner rotations of
Eq.~(\ref{eqmced5}) can be seen as conditional logical operators,
which rotate the spin a certain angle depending on the value of
the momentum. Thus, a Lorentz transformation will modify in
general the entanglement between momentum and spin of each
individual electron. We distinguish the following three cases.

i) \textit{Product state in all variables.} In this case,
\begin{equation}
g_{\lambda\sigma}(\mathbf{p},\mathbf{q}) =
g_1(\mathbf{p})g_2(\mathbf{q})|\lambda\rangle|\sigma\rangle ,
\end{equation}
and the entanglement at the rest reference frame is zero. Under a
boost, the Wigner rotations of Eq.~(\ref{eqmced5}) entangle the
momentum of each fermion with its spin, and thus the entanglement
momentum-spin grows~\cite{PST02}.

ii) \textit{Entangled state spin-spin and/or momentum-momentum.}
We consider now a state
\begin{equation}
g_{\lambda\sigma}(\mathbf{p},\mathbf{q})=f(\mathbf{p} ,
\mathbf{q})|\phi\rangle
\end{equation}
with $|\phi\rangle$ an arbitrary state of the spins, and
$f(\mathbf{p},\mathbf{q})$ an arbitrary state of the momenta. In
this case, a Lorentz boost will entangle in general each spin with
its corresponding momentum, and a careful analysis shows that the
spin-spin entanglement never grows~\cite{GA02}. Of course, by
applying the reversed boost the entanglement momentum-spin would be
transferred back to the spin-spin one, and thus the latter would
grow. This particular case shows that, for the state we considered
in Sec. \ref{ges}, given by Eqs.~(\ref{eqsges1}), (\ref{eqsges1bis})
and (\ref{eqsges3}), the entanglement could not be transferred from
momenta into spins via Lorentz transformations. Thus, the dynamical
approach would be here more suitable.

iii) \textit{Entangled state momentum-spin.} According to the
previous theorem, the momentum-spin entanglement may be lowered,
transferring part of the correlations to the spins, or increased,
taking some part of the correlations from them. To our knowledge,
there is not a similar result for the momentum, that is, whether
the momentum entanglement can be preserved under boosts, or it
suffers decoherence similarly to the spins, and part of it is
transferred to the momentum-spin part. This is a very interesting
question, which we will treat more deeply in future works.


\begin{thebibliography}{24}


\bibitem{epr} A. Einstein, B. Podolsky, N. Rosen, Phys. Rev. {\bf 47}, 777 (1935).

\bibitem{Bell64} J.~S. Bell, Physics {\bf 1}, 195 (1964).

\bibitem{Bennett93} C. H. Bennett, G. Brassard, C. Cr\'epeau, R. Jozsa, A. Peres, and
W. K. Wootters, Phys. Rev. Lett. {\bf 70}, 1895 (1993).

\bibitem{QuantumComm} L. K. Grover, quant-ph/9704012; J. I. Cirac, A. K. Ekert,
S. F. Huelga, and C. Macchiavello, Phys. Rev. A {\bf 59}, 4249
(1999).

\bibitem{Ekert91} A. K. Ekert, Phys. Rev. Lett. {\bf 67}, 661 (1991).

\bibitem{NielsenChuang} M.~A. Nielsen and I.~L. Chuang,
{\it Quantum Computation and Quantum Information} (Cambridge
University Press, Cambridge, England, 2000).

\bibitem{MiguelAngel} A. Galindo and M. A. Mart\'{\i}n-Delgado, Rev.
Mod. Phys. {\bf 74}, 347 (2002).

\bibitem{Englert} B.-G. Englert and K. W\'odkiewicz, Int. J. of
Quant. Inf. {\bf Vol. 1}, No. 2, 153 (2003).

\bibitem{C97} M. Czachor, Phys. Rev. A, {\bf 55}, 72 (1997).

\bibitem{PST02} A. Peres, P.~F. Scudo, and D.~R. Terno,
Phys. Rev. Lett. {\bf 88}, 230402 (2002).

\bibitem{AM02} P.~M. Alsing and G.~J. Milburn,
Quantum Inf. Comput. {\bf 2}, 487 (2002).

\bibitem{GA02} R.~M. Gingrich and C. Adami,
Phys. Rev. Lett. {\bf 89}, 270402 (2002).

\bibitem{GBA03} R.~M. Gingrich, A.~J. Bergou,
and C. Adami, Phys. Rev. A {\bf 68}, 042102 (2003).

\bibitem{PS03} J. Pachos and E. Solano, Quantum Inf. Comput. {\bf 3},
115 (2003).

\bibitem{TU03} H. Terashima and M. Ueda, Quantum Inf. Comput. {\bf 3},
224 (2003).

\bibitem{ALM03} D. Ahn, H. J. Lee, Y. H. Moon, and S. W. Hwang,
Phys. Rev. A {\bf 67}, 012103 (2003).

\bibitem{PT04} A. Peres and D. R. Terno, Rev. Mod. Phys. {\bf 76}, 93 (2004).

\bibitem{MY04} E. B. Manoukian and N. Yongram, Eur. Phys. J. D {\bf 31}, 137 (2004).

\bibitem{qedentang} R. Grobe, K. Rz\c{a}\.{z}ewski, and J. H. Eberly,
J. Phys. B: At. Mol. Opt. Phys. {\bf 27}, L503 (1994).

\bibitem{bethe} E. E. Salpeter and H. E. Bethe, Phys. Rev. {\bf 84}, 1232 (1951).

\bibitem{gellmann} M. Gell-Mann and F. Low, Phys. Rev. {\bf 84}, 350 (1951).

\bibitem{lljl} L. Lamata and J. Le\'on, J. Opt. B: Quantum Semiclass. Opt. {\bf 7},
224 (2005).

\bibitem{entanglefermion1} J. Schliemann, J. I. Cirac, M. Ku\'s, M. Lewenstein,
and D. Loss, Phys. Rev. A {\bf 64}, 022303 (2001).

\bibitem{ESB02} K. Eckert, J. Schliemann, D. Bru\ss, and M. Lewenstein,
Ann. Phys. {\bf 299}, 88 (2002).

\bibitem{entanglefermion2} G. Ghirardi and L. Marinatto, Fortschr.
Phys. {\bf 52}, 1045 (2004).

\bibitem{moshinsky} M. Moshinsky, Phys. Rev. {\bf 88}, 625 (1952).

\bibitem{grads} I. S. Gradshteyn and I. M. Ryzhik, {\it Table of Integrals,
Series, and Products} (Academic Press, Inc., Orlando, 1980),
Equation 7.374.8.

\bibitem{schmidtdisc2} A. Ekert and P. L. Knight, Am. J. Phys. {\bf 63}, 415 (1995).

\bibitem{eberly1} C. K. Law, I. A. Walmsley, and J. H. Eberly,
Phys. Rev. Lett. {\bf 84}, 5304 (2000).

\bibitem{majoriz} M. A. Nielsen, Phys. Rev. Lett. {\bf 83}, 436
(1999).

\end{thebibliography}
\end{document}